\documentclass[english,aps, prl, reprint, showpacs]{revtex4-1}
\usepackage[T1]{fontenc}
\usepackage[latin9]{inputenc}
\usepackage{color}
\usepackage{units}
\usepackage{ifsym}
\usepackage{amsmath}
\usepackage{wasysym}
\usepackage{graphicx}

\makeatletter

 \@ifundefined{textcolor}{}
 {%
   \definecolor{BLACK}{gray}{0}
   \definecolor{WHITE}{gray}{1}
   \definecolor{RED}{rgb}{1,0,0}
   \definecolor{GREEN}{rgb}{0,1,0}
   \definecolor{BLUE}{rgb}{0,0,1}
   \definecolor{CYAN}{cmyk}{1,0,0,0}
   \definecolor{MAGENTA}{cmyk}{0,1,0,0}
   \definecolor{YELLOW}{cmyk}{0,0,1,0}
 }

\makeatother

\usepackage{babel}
\begin{document}

\title{Creation of prompt and thin-sheet splashing by varying surface roughness or increasing air pressure}

\author{Andrzej Latka}

\email{alatka@uchicago.edu}

\selectlanguage{english}%

\affiliation{The James Franck Institute and Department of Physics, The University
of Chicago, Chicago, Illinois 60637, USA}

\author{Ariana Strandburg-Peshkin}

\affiliation{The James Franck Institute and Department of Physics, The University
of Chicago, Chicago, Illinois 60637, USA}

\author{Michelle M. Driscoll}

\affiliation{The James Franck Institute and Department of Physics, The University
of Chicago, Chicago, Illinois 60637, USA}

\author{Cacey S. Stevens}

\affiliation{The James Franck Institute and Department of Physics, The University
of Chicago, Chicago, Illinois 60637, USA}

\author{Sidney R. Nagel}

\affiliation{The James Franck Institute and Department of Physics, The University
of Chicago, Chicago, Illinois 60637, USA}
\begin{abstract}
A liquid drop impacting a solid surface may splash either by emitting a thin liquid sheet that subsequently breaks apart or by promptly ejecting droplets from the advancing liquid-solid contact line. Using high-speed imaging, we show that surface roughness and air pressure influence both mechanisms. Roughness \emph{inhibits} thin-sheet formation even though it also \emph{increases} prompt splashing at the advancing contact line. If the air pressure is lowered, droplet ejection is suppressed not only during thin-sheet formation but for prompt splashing as well.
\end{abstract}

\pacs{47.20.Cq, 47.20.Gv, 47.20.Ma, 47.55.D-}

\maketitle

Will a drop hitting a dry surface splash? Different criteria \cite{Stow1981,Range1998,Mundo1995,Rioboo2001,Yarin1995} have been proposed to predict when such a drop will splash by comparing the roughness of the solid surface with hydrodynamic length scales, which depend on parameters such as the drop velocity, radius, viscosity and surface tension. Several years ago Xu et al.\ \cite{Xu2005,Xu2007} found that these criteria ignore a crucial parameter: the ambient gas pressure, $P$. When a drop splashes on a smooth surface it spreads smoothly forming a lamella before ejecting a thin sheet that subsequently breaks up into secondary droplets. As $P$ is reduced below a threshold pressure, the drop no longer splashes~\cite{Xu2005,Xu2007,Xu2007a,Driscoll2010,bang2011assessment}. On the other hand, when splashing occurs on a rough surface, no thin sheet is formed and droplets are ejected directly from the advancing liquid-substrate contact line via a \textquotedbl{}prompt\textquotedbl{} splash \cite{Stow1981,Mundo1995,Range1998,Xu2007a,Rioboo2001}. 

It has been suggested that thin-sheet splashes depend on air pressure while prompt splashes do not and depend only on surface roughness~\cite{Xu2007a}. Here we show that the situation is more complex in that both types of splashing depend, albeit in opposite ways, on surface roughness. In particular, we observe four distinct regimes. In agreement with earlier results \cite{Rioboo2001}, we observe a thin-sheet splash on very smooth surfaces and a prompt splash on very rough ones. However, at intermediate roughness, we identify two new regimes: at low viscosities both prompt and thin-sheet splashes occur during a single impact, while at high viscosities neither splash is formed. In addition, as found for thin-sheet splashing~\cite{Xu2005}, we find that a drop deposits smoothly on a rough surface if $P$ is low enough. Clearly, the role of both air pressure and substrate roughness must be considered in all cases. 


The experiments were conducted with silicone oil (PDMS, Clearco Products) with kinematic viscosity $\nu$ ranging from $5$ cSt to $14.4$ cSt and surface tension $\sigma$ between $19.7$ dyn/cm and $20.8$ dyn/cm. The basic results were replicated using water/glycerin mixtures with a similar viscosity range but higher surface tension: $\sigma$=$67$ dyn/cm. Low-viscosity impacts were studied with ethanol. Drops with reproducible diameter $D$=$3.1$ mm were produced using a syringe pump (Razel Scientific, Model R99-E) and released in a chamber from a height above a substrate. This height set the impact velocity $u_{0}$ which was varied between $2.7$ m/s and $4.1$ m/s. These parameters determine the Reynolds number Re=$D u_{0}$/$\nu$ giving the ratio of inertial to viscous forces, and the Weber number We=$\rho D u_{0}^2$/$\sigma$ giving the ratio of inertial to surface tension forces. Here we consider the regime $580$<Re<$7100$ and $390$<We<$2400$. The chamber could be evacuated down to $P$=$5$ kPa. Impacts were recorded with a high-speed camera (Phantom v12, Vision Research) at rates up to $44,000$ fps.

Three types of rough surfaces were used: commercially prepared sandpaper consisting of aluminum-oxide particles, glass slides (Fisherbrand Microscope Slides) etched with ammonium bifluoride (Armour Etch cream) for different times, and acrylic plates roughened with sandpapers of varying grit.  The surface height $h(r_n)$ was measured at $N$ equally spaced positions $r_n$ using atomic force microscopy (Asylum MFP-3D AFM) on several square patches of side $L$=$75$ $\mu m$. From these profiles we determined the root-mean-square roughness $R_{rms}$, which varies from $0.005$ $\mu m$ to $2.84$ $\mu m$ and the roughness power spectrum: $H\left(k\right)=\left\langle\frac{1}{N}\left|\underset{n=0}{\overset{N-1}{\sum}}h\left(r_{n}\right)e^{i\mathbf{k \cdot r_n}}\right|^{2} \right\rangle $. The brackets indicate an average over patches and over directions. Each surface was used only once to prevent contamination from previous splashes. 

\begin{figure*}[t]
\includegraphics[width=1\textwidth]{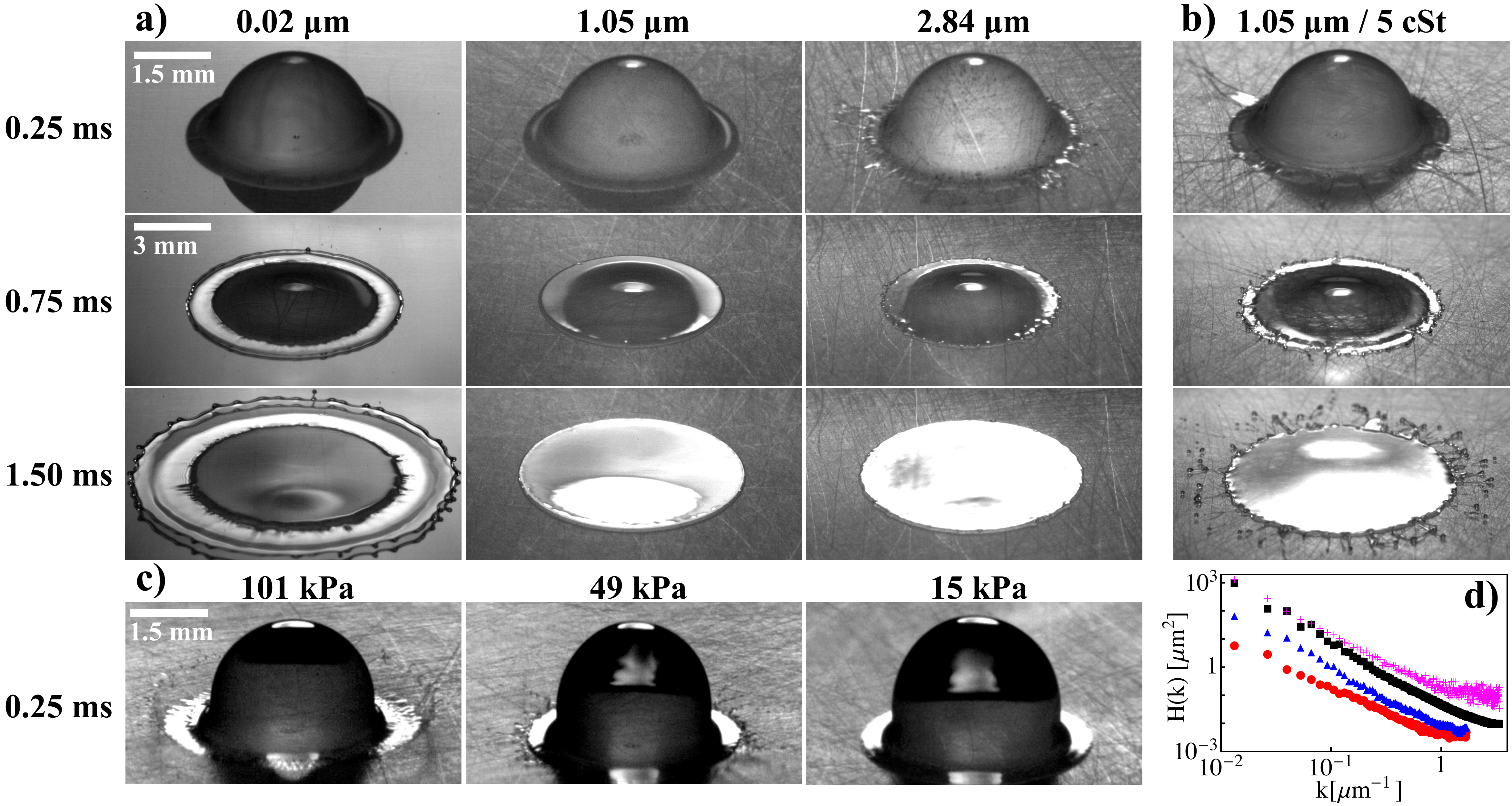}

\caption[]{\label{fig:stills}
a) Images of a $10$ cSt silicone oil drop with $u_{0}$=$4.1$ m/s (Re=$1300$, We=$2400$) impacting acrylic surfaces with root-mean-square roughness $0.02$ $\mu m$ (left), $1.05$ $\mu m$ (middle), and $2.84$ $\mu m$ (right) at atmospheric pressure. The images were taken $0.25$ ms (top), $0.75$ ms (middle) and $1.50$ ms (bottom) after impact. The smoothest surface produces a thin-sheet splash, the roughest produces a prompt splash, while the intermediate roughness surface completely suppresses splash formation. 
b) Impact of a $5$ cSt silicone oil drop (Re=$2500$, We=$2400$) on acrylic with $R_{rms}$=$1.05$ $\mu m$.  A prompt splash is formed, followed by thin-sheet ejection and splash. 
c) An ethanol drop ($\nu$=$1.36$ cSt) at $u_{0}$=$3.1$ m/s (Re=$7100$, We=$1050$) shown $0.25$ ms after impact on rough acrylic ($R_{rms}$=$1.05$ $\mu m$):  at $P$=$101$ kPa (left), $49$ kPa (middle) and $15$ kPa (right). Prompt splashing is suppressed at lower gas pressure.  All images at $t=0.25$ ms are magnified to show prompt splashing.
d) Roughness power spectrum, $H(k)$, for rough acrylic with $R_{rms}$=$0.34$ $\mu m$ (\includegraphics[height=6pt]{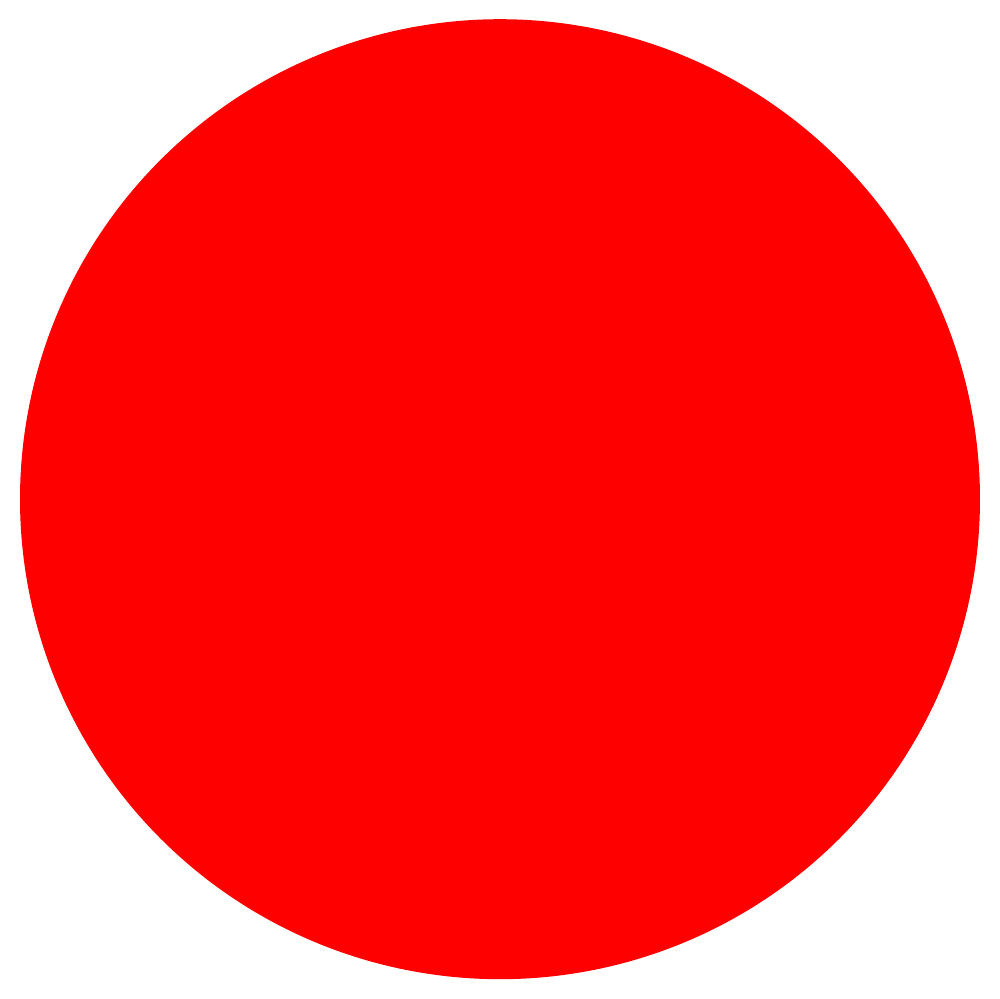}), $R_{rms}$=$1.05$ $\mu m$ (\includegraphics[height=6pt]{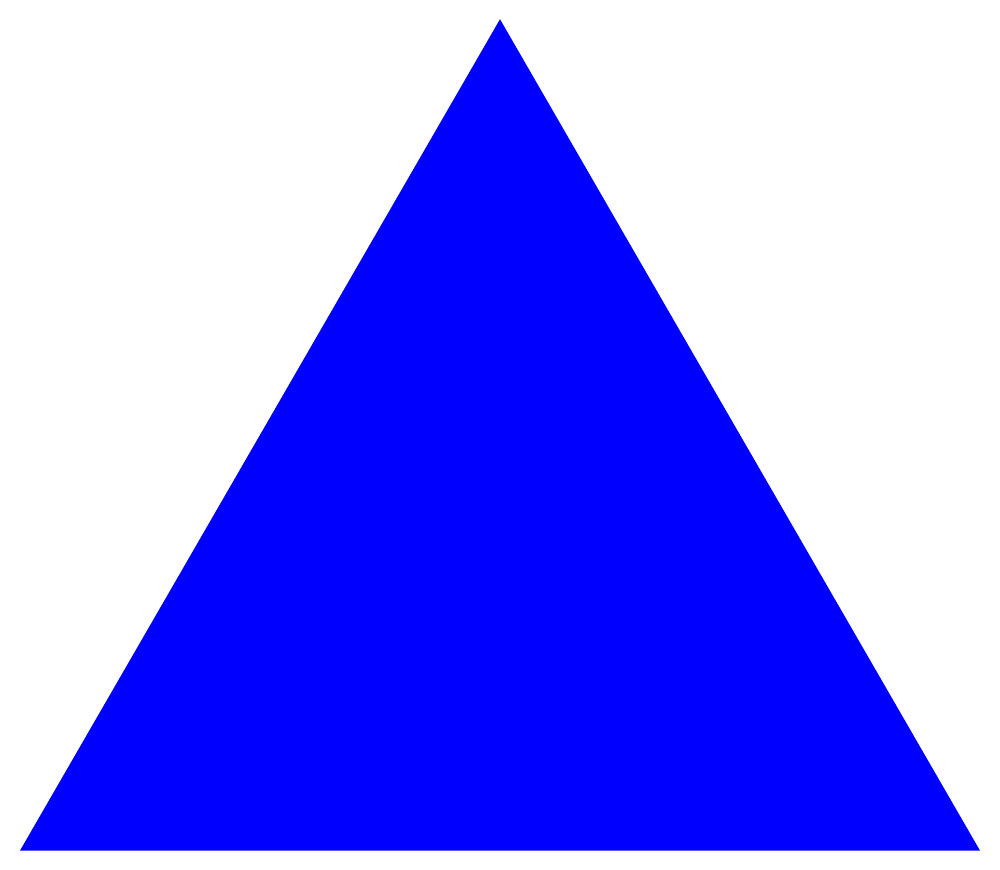}) and $R_{rms}$=$2.84$ $\mu m$ (\includegraphics[height=6pt]{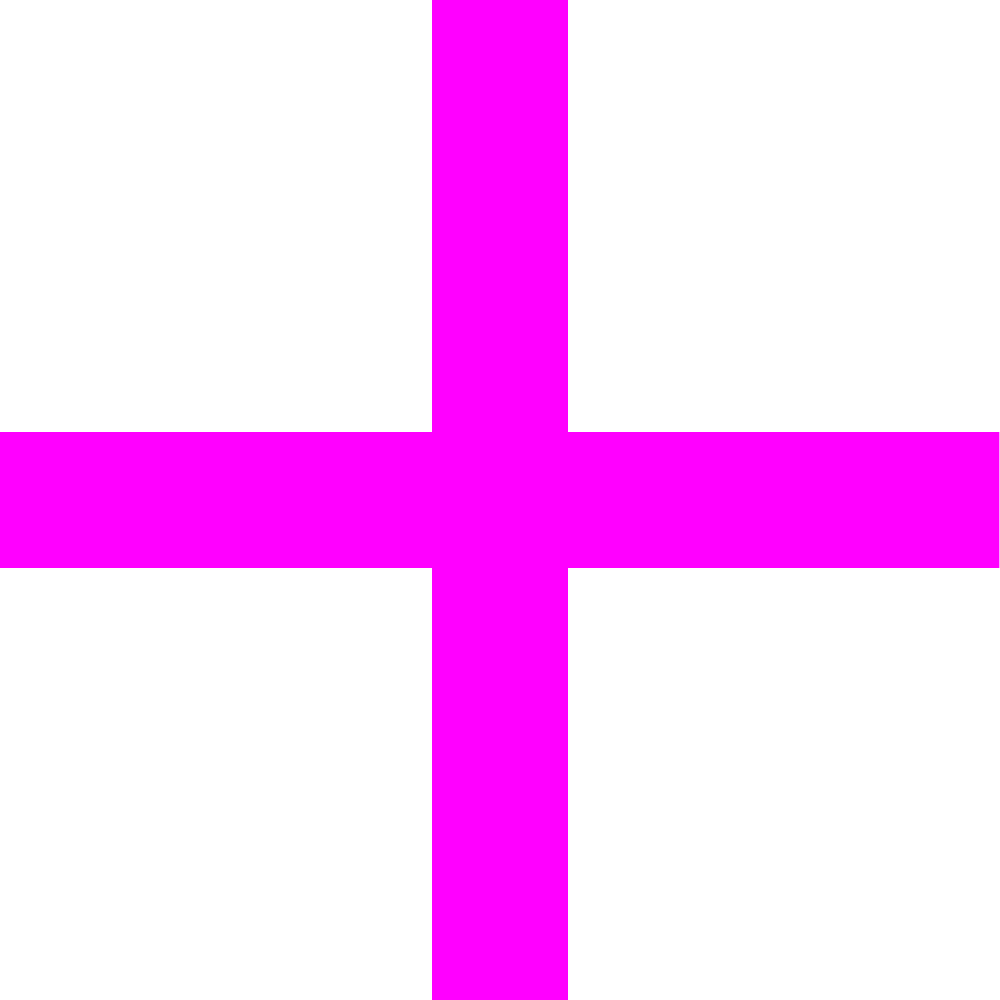}) and of sandpaper with $R_{rms}$=$2.47$ $\mu m$ (\includegraphics[height=6pt]{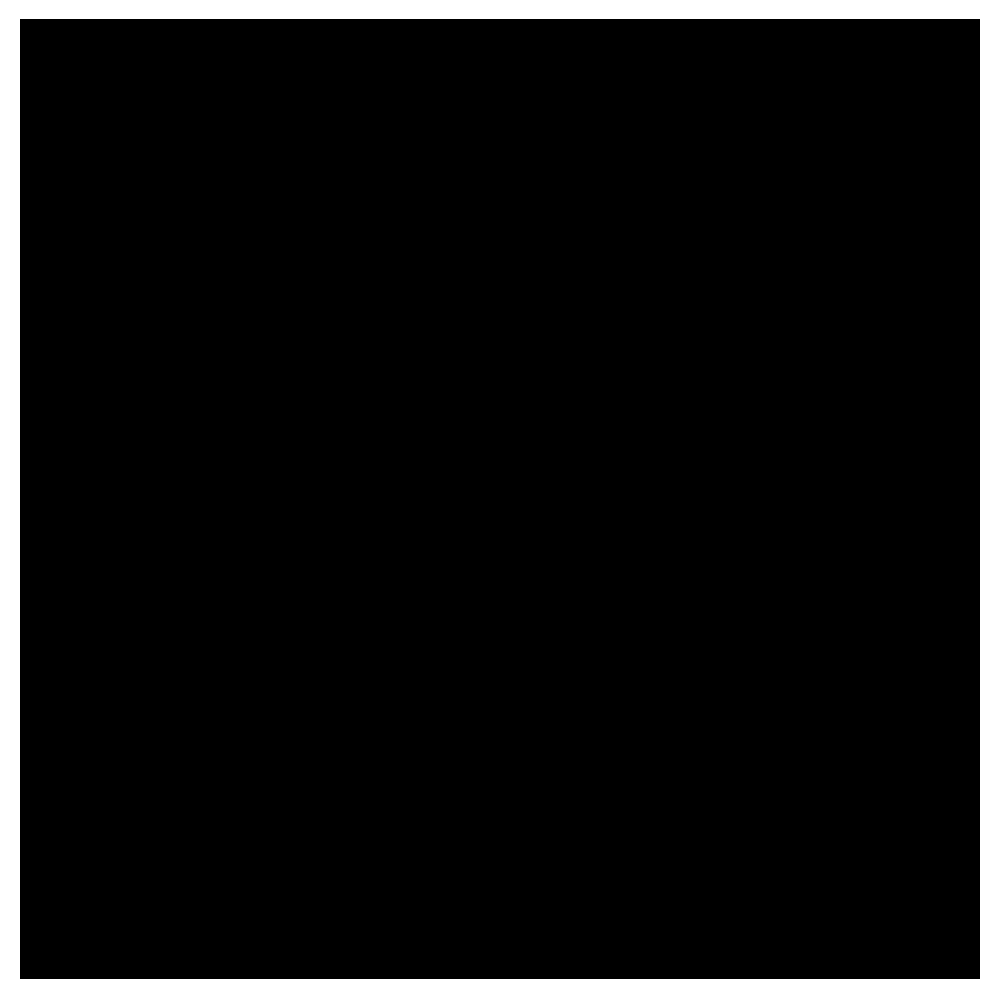}). The surfaces have the same qualitative shape of $H(k)$. }
\end{figure*}


Previous studies of water and ethanol droplets \cite{Stow1981,Range1998,Xu2007a} concluded that increasing surface roughness promotes splashing.  Fig.\ \ref{fig:stills} shows that when one considers more viscous liquids, increasing the surface roughness can in fact completely suppress it. The left column of Fig.\ \ref{fig:stills}\emph{a} shows a $10$ cSt drop creating a thin-sheet splash on a smooth surface with $R_{rms}$=$0.02$ $\mu m$. The top image shows the lamella spreading before thin-sheet ejection. The middle image shows the ejected thin sheet, and the bottom image shows the thin sheet during break-up. The middle column in Fig.\ \ref{fig:stills}\emph{a} shows an identical drop under the same conditions impacting an intermediate roughness acrylic plate with $R_{rms}$=$1.05$ $\mu m$. As the drop spreads, the thin sheet fails to emerge: a small amount of roughness completely suppresses splashing. The right column of Fig.\ \ref{fig:stills}\emph{a} shows drop impact on an even rougher surface with $R_{rms}$=$2.84$ $\mu m$. Immediately after impact liquid is ejected from the lamella at the spreading liquid-solid contact line to form a prompt splash. At a lower viscosity the drop splashes via both mechanisms. Fig.\ \ref{fig:stills}\emph{b} shows that a $5$ cSt drop initially undergoes a prompt splash (compare \ref{fig:stills}\emph{b} top and \ref{fig:stills}\emph{a} top right). Later, however, it also ejects a thin sheet  (\ref{fig:stills}\emph{b} middle), which then disintegrates into droplets (\ref{fig:stills}\emph{b} bottom).

Fig.\ \ref{fig:stills}\emph{c} shows the impact of a low-viscosity liquid drop, ethanol, on acrylic ($\nu$=$1.36$ cSt, $R_{rms}$=$1.05$ $\mu m$). At atmospheric pressure, we see a prominent prompt splash with many droplets ejected at large angles from the spreading contact line. On decreasing the pressure to $P=49$ kPa, droplets become much fewer in number and are ejected almost parallel to the surface. Below $15$ kPa prompt splashing is completely suppressed.  

Fig.\ \ref{fig:stills}\emph{d} shows examples of the roughness power spectrum, $H(k)$, for the surfaces that we used. Nilsson et al. \cite{nilsson2010novel} found that surfaces of sandpapered Teflon exhibit roughness on many length scales. Fig.\ \ref{fig:stills}\emph{d} shows that our similarly prepared acrylic surfaces also do not feature a characteristic length. However they have a characteristic shape for $H(k)$ with a coefficient determined by $R_{rms}$.


\begin{figure}
\includegraphics[width=1\columnwidth]{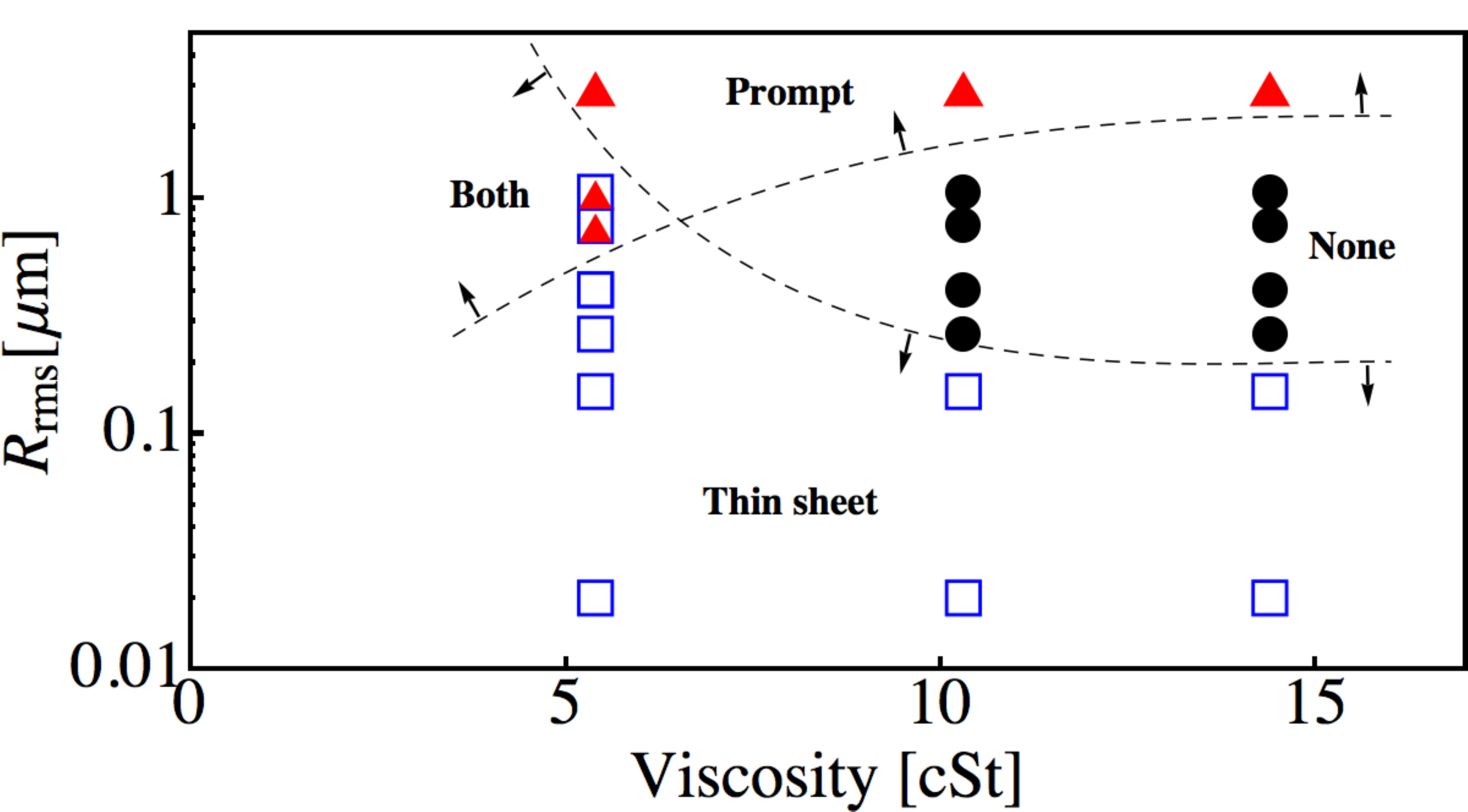}

\caption[]{\label{fig:acrylic_phase_diagram}Splashing phase diagram for silicone oil drops impinging on an acrylic surface at $u_{0}$=$3.4$ m/s ($730$<Re<$1900$, We$\approx$1650) and atmospheric pressure. At low roughnesses, the impact results in a thin-sheet splash (\includegraphics[height=6pt]{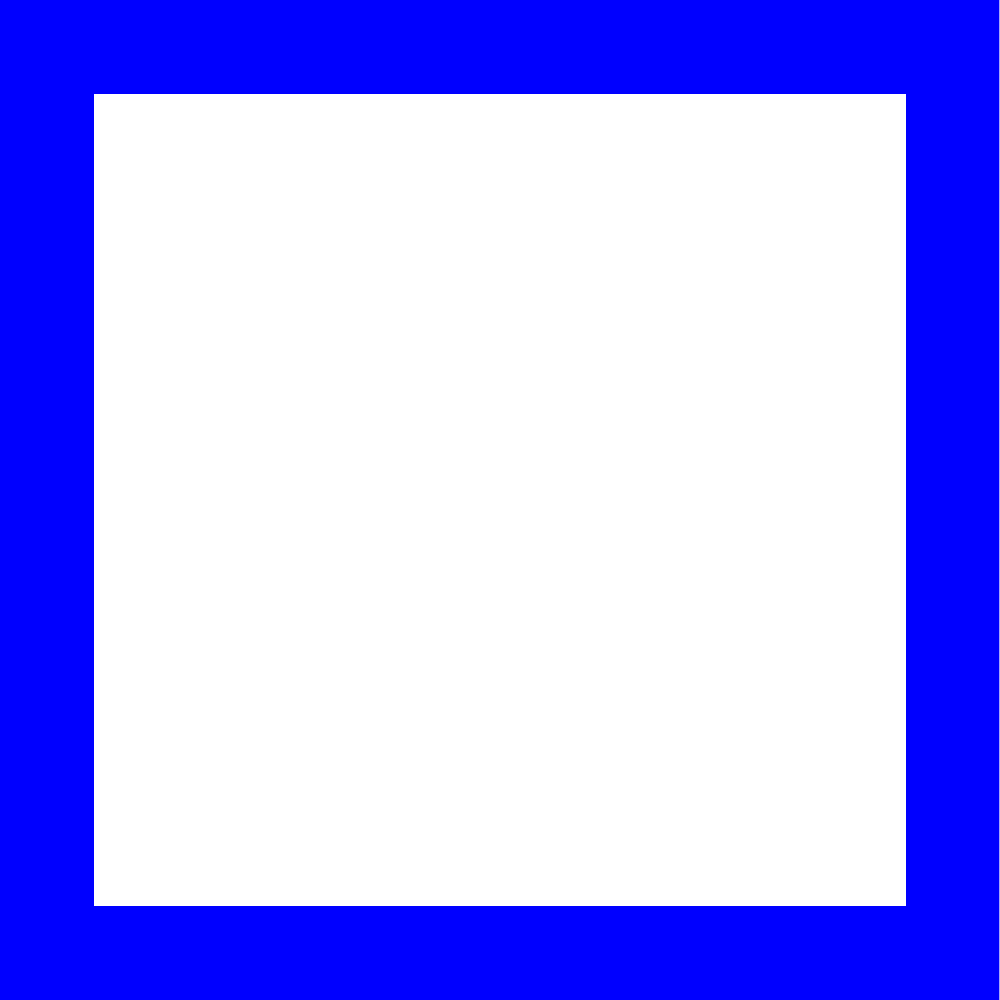}) and at high roughnesses - a prompt splash (\includegraphics[height=6pt]{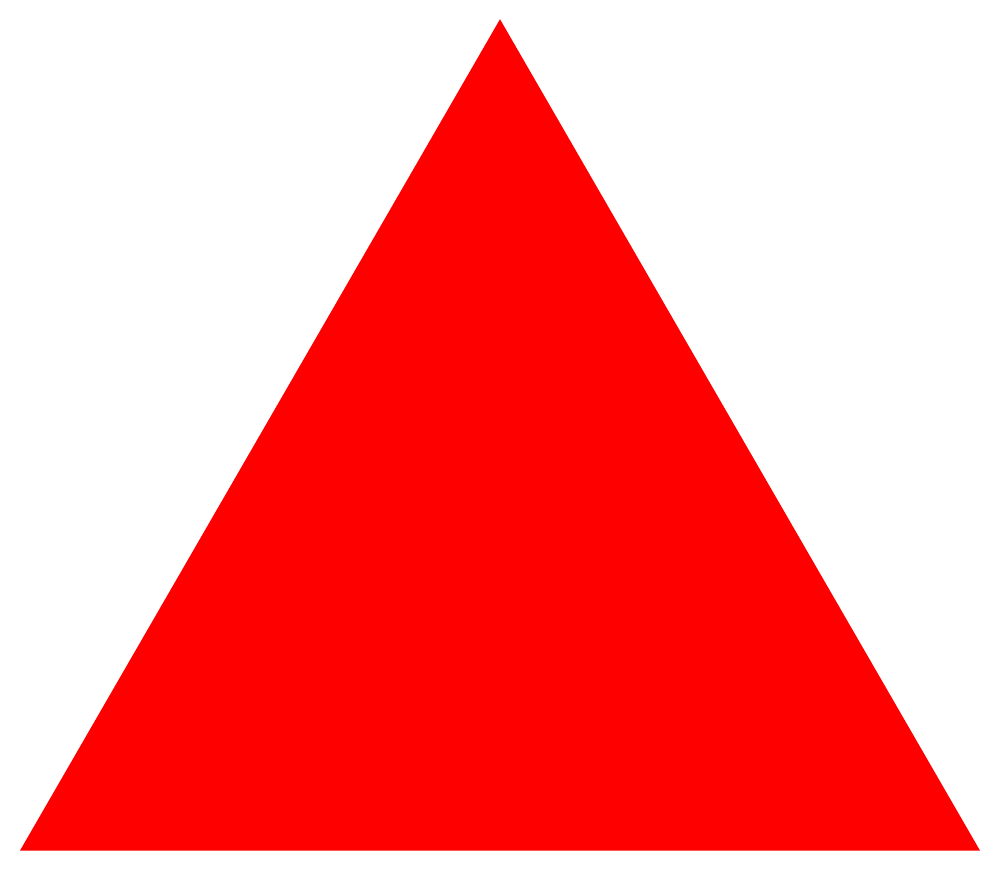}). At intermediate roughnesses two outcomes are possible. At lower viscosities, a prompt splash is followed by a thin-sheet splash in the same impact event (\includegraphics[height=6pt]{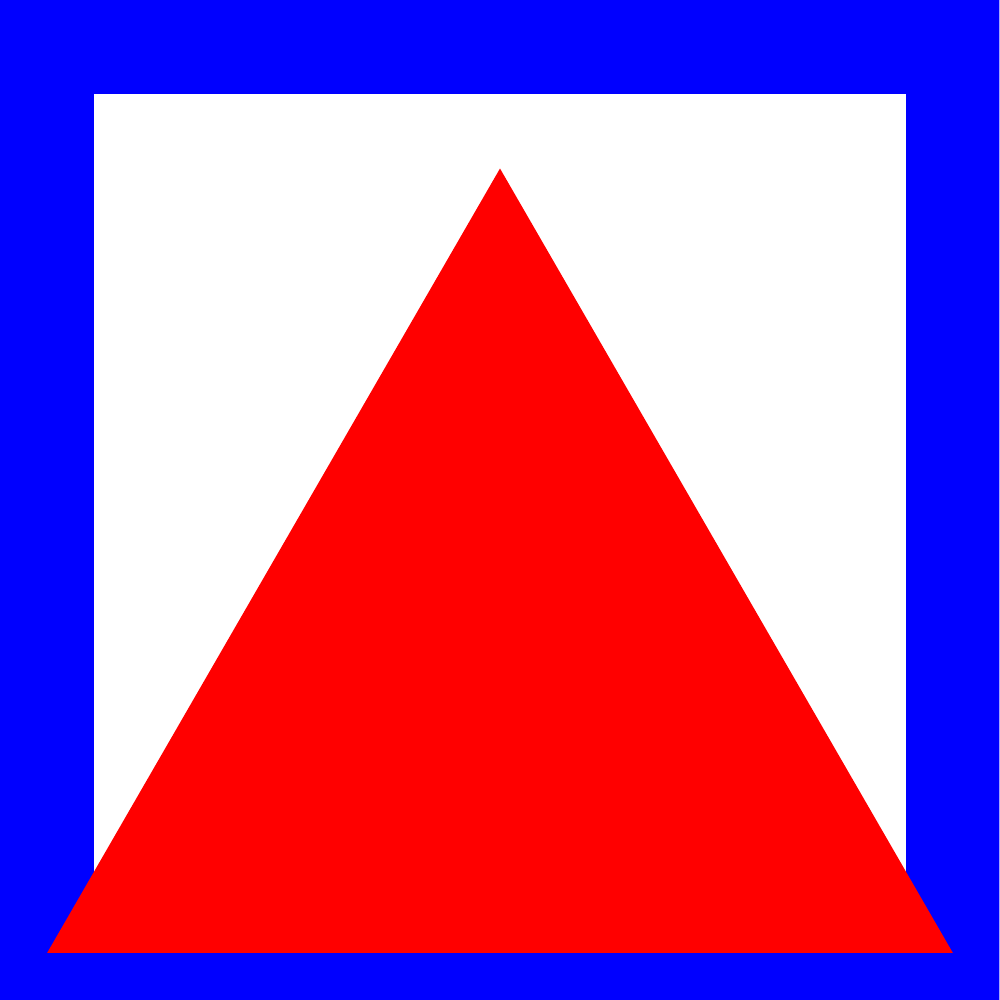}). For higher viscosities, splashing is suppressed completely (\includegraphics[height=6pt]{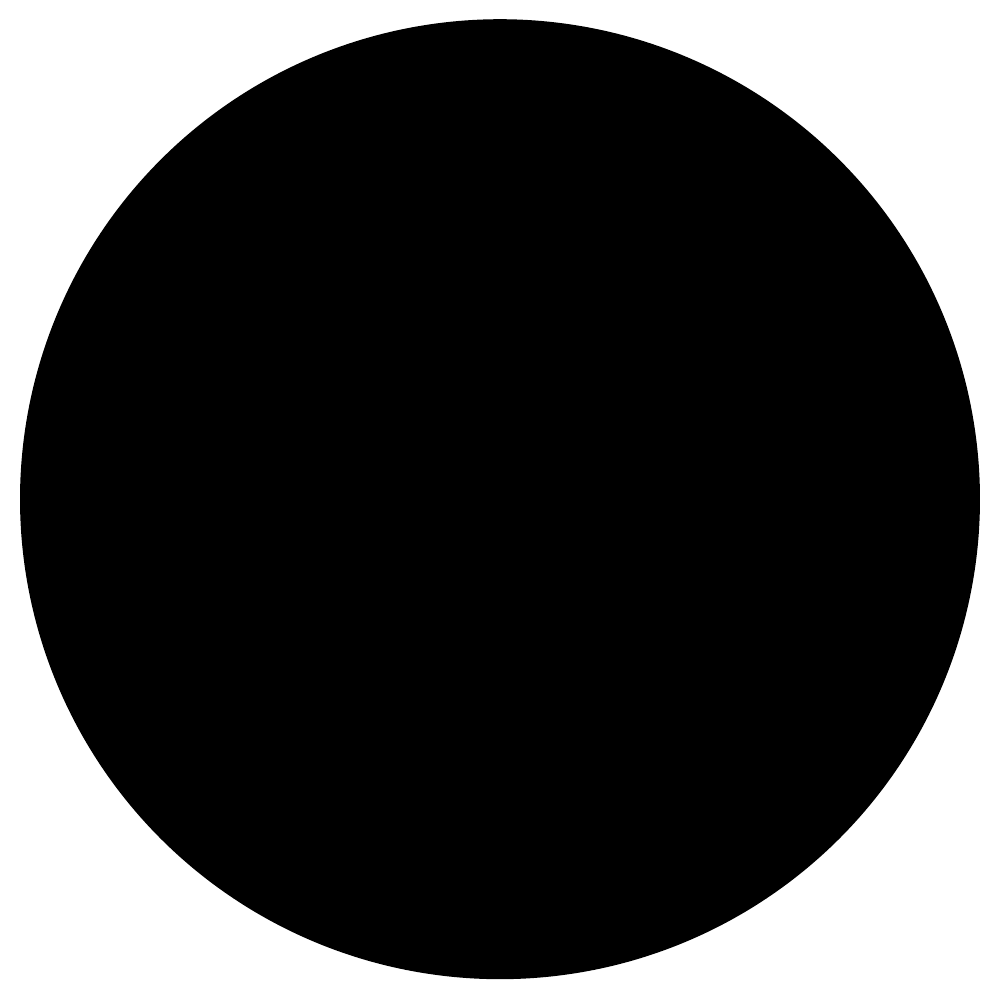}): one can suppress splashing by \emph{increasing} the roughness of the surface. The dashed lines separating different splashing regimes are guides to the eye, and shift in the direction given by the arrows as air pressure is reduced. }

\end{figure}

The behavior at atmospheric pressure is summarized in the form of a phase diagram in Fig.\ \ref{fig:acrylic_phase_diagram}. Surface roughness affects prompt splashing and thin-sheet splashing in opposite ways. Consider first the thin-sheet splash. For this splash to occur, the roughness of the substrate must be below a threshold value. For example, a $5.5$ cSt drop will undergo a thin-sheet splash only for $R_{rms}$ < $2.84$ $\mu m$. The threshold roughness decreases as the drop viscosity is increased. Thus, a $10.3$ cSt drop undergoes thin-sheet splashing only when $R_{rms}$ < $0.26$ $\mu m$. 

On the other hand, the prompt splash only occurs above a threshold roughness that \emph{increases} with viscosity. Consequently, both splashes take place for $5.5$ cSt drops at intermediate roughnesses, because the roughness necessary to create a prompt splash is insufficient to prevent a thin-sheet splash. Increasing the viscosity leads to a regime in which no splashing occurs: the roughness prevents a thin-sheet splash but is insufficient to produce a prompt splash. We obtained qualitatively similar phase diagrams, consisting of four regions, on the three different types of rough surfaces described above using either water/glycerin mixtures or silicone oil.


\begin{figure}
\includegraphics[width=1\columnwidth]{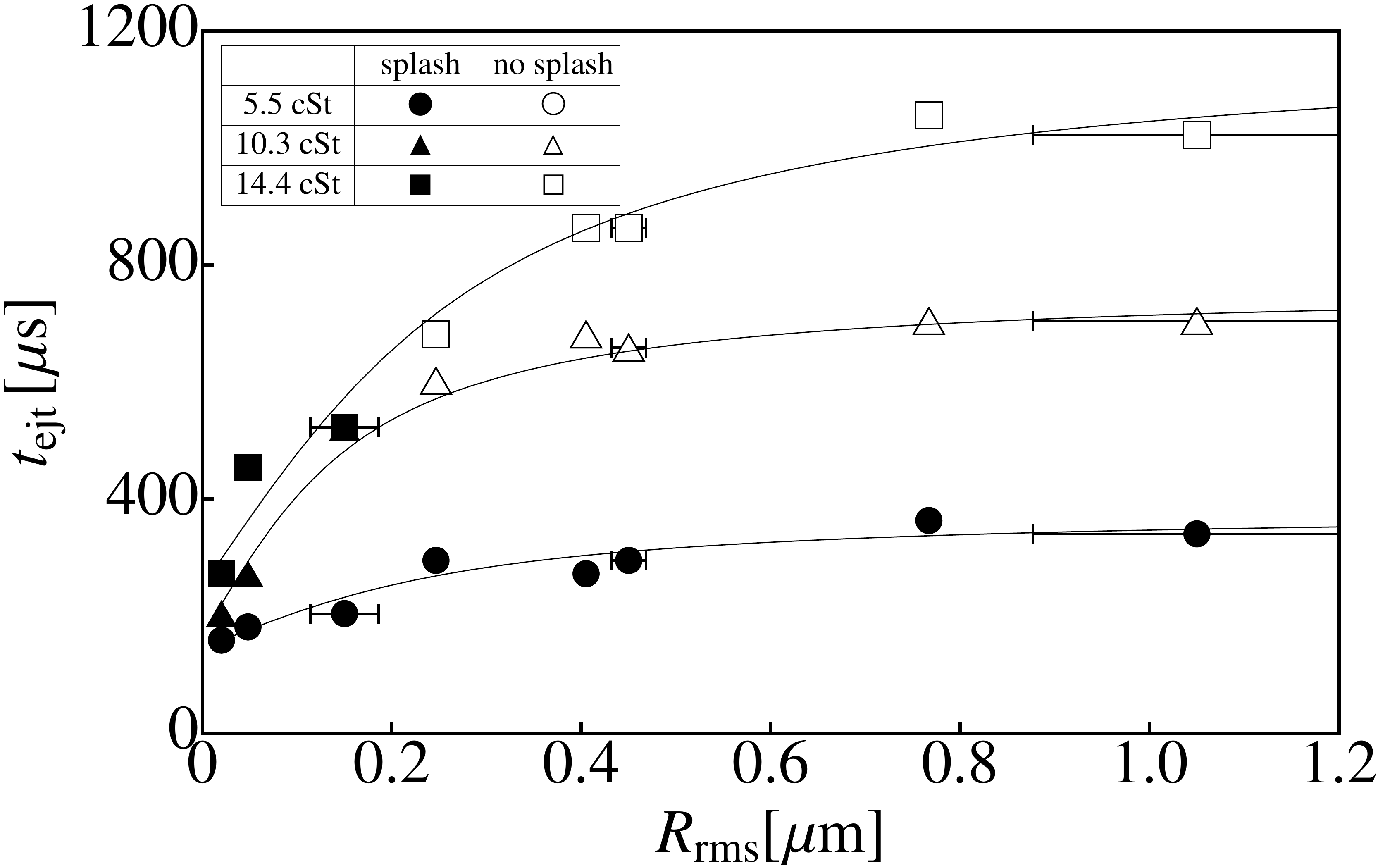}

\caption[]{\label{fig:ejection_times} Thin-sheet ejection times vs. surface roughness for silicone oil drops of viscosity $5.5$ cSt (\includegraphics[height=6pt]{BlackCircle.pdf}), $10.3$ cSt (\includegraphics[height=6pt]{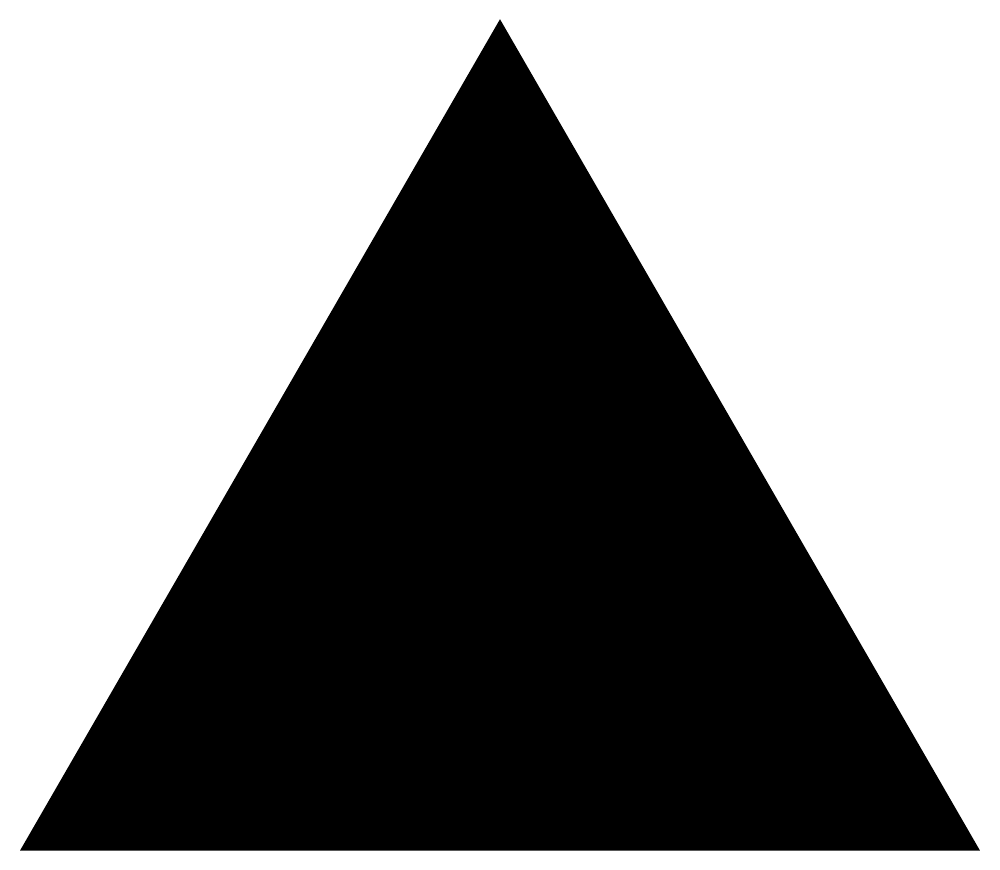}), and $14.4$ cSt (\includegraphics[height=6pt]{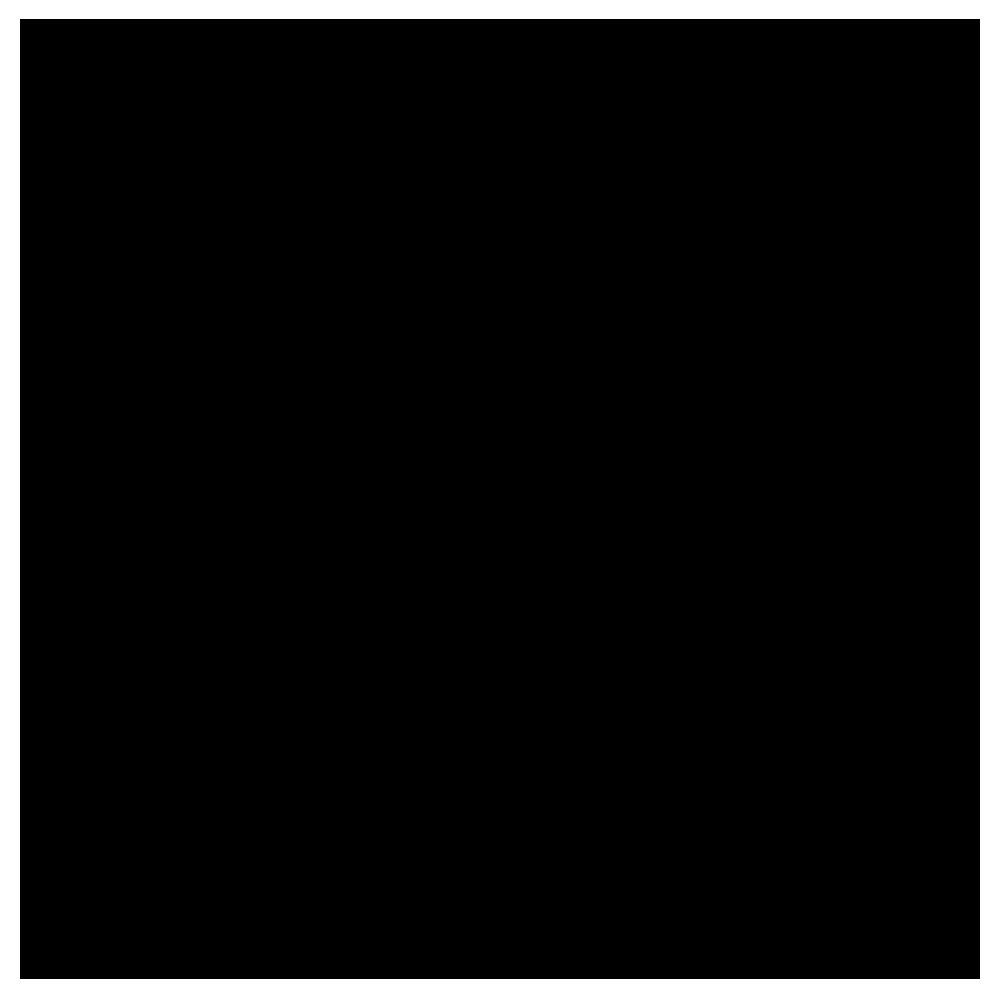}) impacting an acrylic surface at $u_{0}$=$3.4$ m/s ($730$<Re<$1900$, We$\approx$1650). As $R_{rms}$ increases, $t_{ejt}$ increases. For the higher viscosities and $R_{rms}$>$0.23$ $\mu m$, the thin sheet no longer breaks apart into smaller droplets (open symbols).}
\end{figure}

The mechanism by which drop impact creates a splash is not understood and different possibilities, such as air entrapment, have been explored. Even for smooth substrates, air is trapped below a drop on impact \cite{chandra1991collision, thoroddsen1998evolution}. This has been studied theoretically \cite{mani2010events, Duchemin2011} and experimentally \cite{driscoll2011ultrafast,kolinski2012skating,de2012dynamics,van2012direct} but found not to be relevant for the splashing of viscous drops \cite{driscoll2011ultrafast}. To gain insight as to the cause of splashing, we must first determine carefully the way a splash is generated. A viscous drop impacting a smooth dry surface forms a thick lamella at time $\nu$/$u_{0}^2$~\cite{mongruel2009early,de2010thickness}. As shown in Fig.\ \ref{fig:stills}\emph{a}, this lamella first spreads smoothly until, at a time $t_{ejt} \gg \nu $/$u_{0}^2$, it ejects a much thinner sheet ~\cite{Driscoll2010}. Fig.\ \ref{fig:ejection_times} shows that $t_{ejt}$ increases as the substrate roughness $R_{rms}$ is increased. This effect grows with increasing viscosity. For example, for a $10.3$ cSt silicone oil drop, $t_{ejt}$ increases more than threefold as $R_{rms}$ is increased from $0.02$ $\mu m$ to $1.05$ $\mu m$. The delay in $t_{ejt}$ decreases the size of the ejected thin sheet so that it may no longer break apart. Therefore, roughness inhibits thin-sheet splashing by delaying $t_{ejt}$. This resembles the effect of reducing the ambient gas pressure \cite{Driscoll2010}.


\begin{figure}
\includegraphics[width=1\columnwidth]{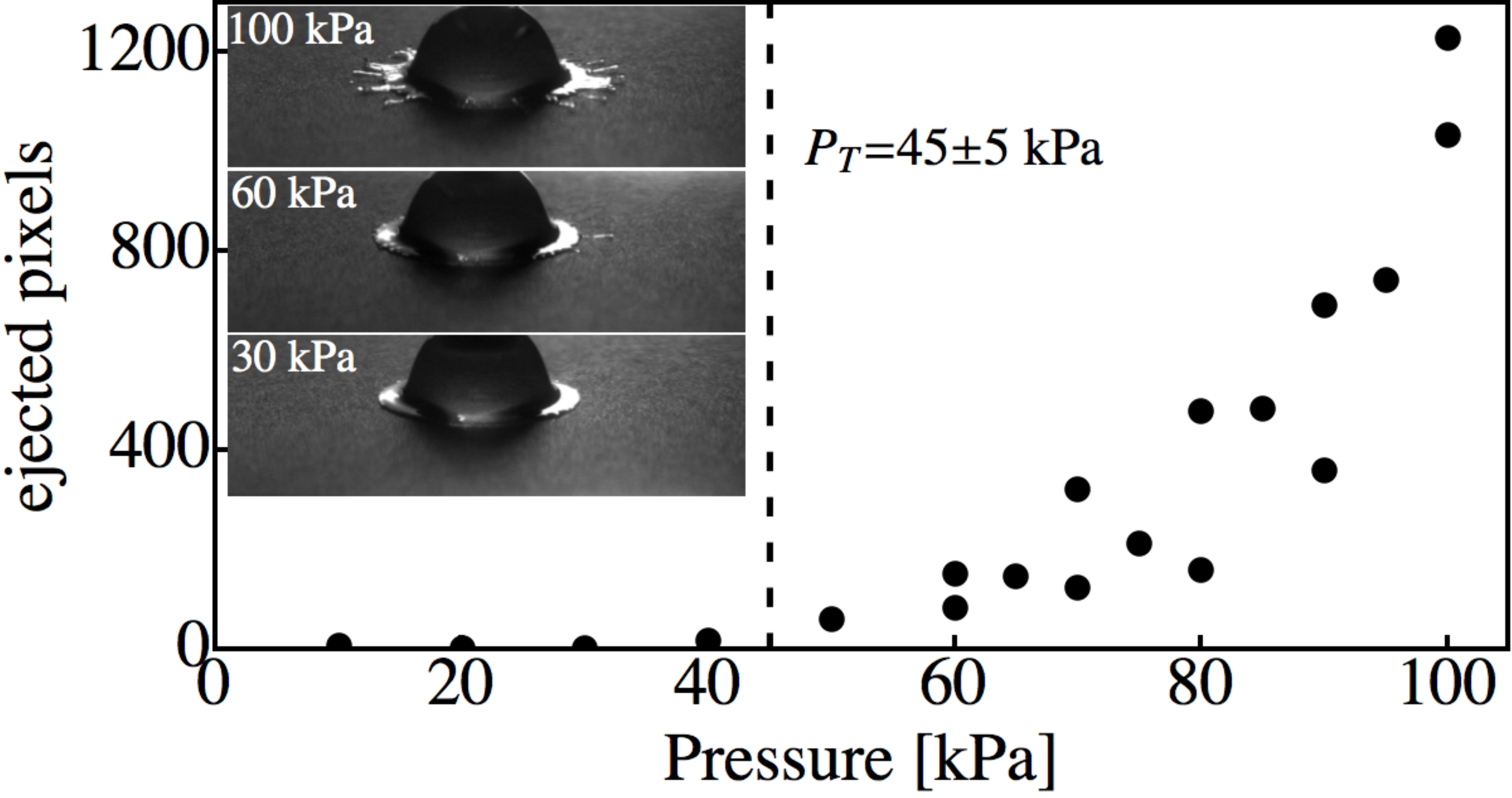}\

\caption{\label{fig:PT_definition}Size of the prompt splash vs. gas pressure for silicone oil drops with viscosity $7.1$ cSt impacting sandpaper with $R_{rms}$=$2.27$ $\mu m$ at $u_{0}$=$2.7$ m/s (Re=$1200$, We=$1050$). Size of splash is obtained from the area in pixels where liquid is not part of the spreading lamella. Each point represents one drop impact and their spread reflects the drop-to-drop fluctuations. As $P$ is decreased, the size of the prompt splash also decreases on average, until at a threshold pressure $P_{T}$=$45\pm5$ kPa splashing disappears. The images show drop impacts $0.34$ ms after impact. At atmospheric pressure many ejected droplets are seen, at $60$ kPa only one is clearly visible, and at $30$ kPa the drop spreads smoothly. }
\end{figure}

\begin{figure}
\includegraphics[width=1\columnwidth]{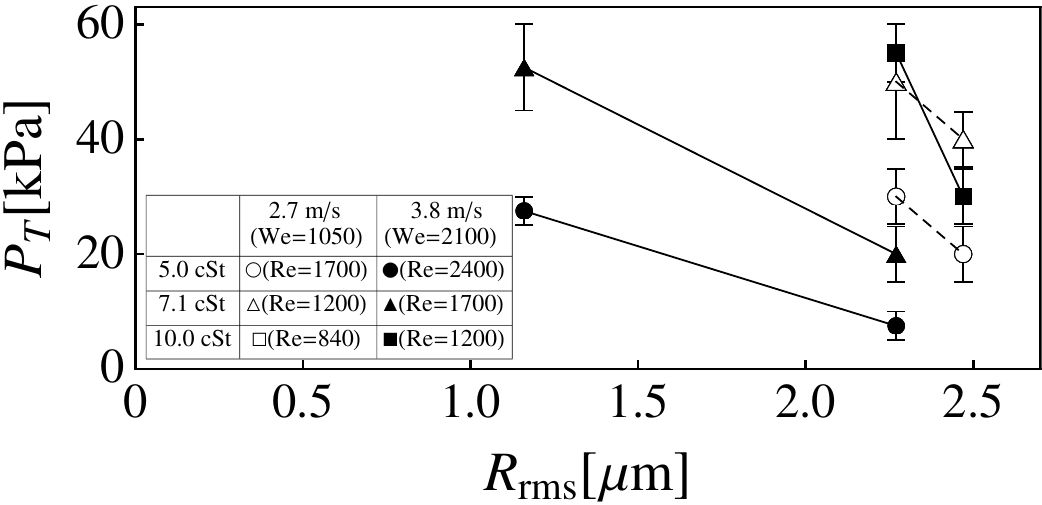}\caption{\label{fig:PT_measured} Threshold pressure of prompt splashing of silicone oil on sandpaper for different values of surface roughness and drop parameters as shown in the table. $R_{rms}$=$1.16,\,2.27,\,2.47$ $\mu m$, liquid viscosity  $\nu$=$5.0,\,7.1,\,10.0$ cSt and impact velocity $u_{0}$=$2.7$, $3.8$ m/s. In each case, increasing the surface roughness and impact velocity decreased the measured threshold pressure, while increasing the liquid viscosity increased it. }
\end{figure}

In contrast, a prompt splash does not involve a thin sheet whose ejection can be delayed. In fact, prompt and thin-sheet splashes are two distinct processes: they depend oppositely on roughness, occur at different times and can take place independently of each other. One would not necessarily expect the ambient air pressure $P$ to affect prompt splashing. Surprisingly, Fig.\ \ref{fig:stills}\emph{c} shows that reducing $P$ inhibits a prompt splash in ethanol - just as it would a thin-sheet splash. This is also true for more viscous drops. Fig.\ \ref{fig:PT_definition} shows that as $P$ decreases, less liquid is ejected until, below a threshold pressure, $P_{T}$, splashing is suppressed completely. $P_{T}$ is a function of $R_{rms}$, $u_{0}$, $\nu$ and $\sigma$. In Fig.\ \ref{fig:PT_measured} we show $P_{T}$ measured from silicone oil impacts on sandpapers of different roughnesses. Increasing $R_{rms}$ or $u_{0}$ decreases $P_{T}$. On the other hand, increasing $\nu$ increases $P_{T}$. 
We have confirmed that prompt splashing is eliminated below a threshold air pressure on all three rough surfaces that we have used (rough acrylic, sandpaper and etched glass) and with different liquids (silicone oil, water-glycerin mixtures and ethanol).


Previous experiments \cite{sivakumar2005spreading, Xu2007, Tsai2010} studied splashing on textured surfaces comprised of micron-sized pillars arranged on a regular lattice. Splashing varied with geometry and was suppressed at low pressures or when the pillars exceeded a certain height.  This was interpreted in terms of air escaping along the linear channels between pillars. Our experiments using surfaces with isotropic, random roughness show trends with $R_{rms}$ and $P$ that are the same on all substrates studied. However, the idea of air escaping along channels cannot be applied to our situation where no such channels exist. 

The inclusion of $R_{rms}$ and $P$ as control parameters is not only important from a practical standpoint, allowing a form of deposition control, but also provides insight into the physical mechanisms of splash formation. Our results extend previous splashing criteria \cite{Stow1981,Range1998,Mundo1995,Rioboo2001,Yarin1995} that were generally determined only at atmospheric pressure and which found a monotonic dependence on roughness. We find a different result. Roughness can suppress thin-sheet splashing while also promoting prompt splashing. Moreover, a sufficiently high air pressure is needed not only for thin-sheet formation ahead of the spreading lamella, but also for the ejection of droplets in a prompt splash directly from the lamella-substrate contact line. The boundaries in the splashing phase diagram, Fig.\ \ref{fig:acrylic_phase_diagram}, shift in the direction of the arrows as pressure is lowered. For sufficiently low $P$ splashing can be suppressed even on very rough surfaces; at high $P$ even relatively smooth surfaces can produce a prompt splash.

It is a formidable task to express the criterion for splashing in terms of dimensionless variables. Previous work revealed several distinct splashing regimes (determined by impact velocity and viscosity) that each depend in different ways on the control parameters~\cite{Xu2005,Xu2007,Driscoll2010}. Moreover, these parameters must include both gas molecular weight and surface roughness, which do not appear in the commonly used numbers such as $Re$ and $We$. Here, we have demonstrated that, in addition, there must be two \emph{separate} criteria: one for thin-sheet and one for prompt splashing.  Our work has shown what additional and unexpected variables must be included in order to produce threshold criteria for both prompt and thin-sheet splashing. Finding the precise form of the splashing criteria in terms of dimensionless variables should be the focus of future work.  

\begin{acknowledgments}
We thank Qiti Guo and Justin Burton. This work was supported by NSF Grants DMR-0652269, REU PHY-1062785 and the MRSEC DMR-0820054. 

\end{acknowledgments}
\bibliographystyle{apsrev4-1}

\end{document}